\documentclass[%
aps,prb,superscriptaddress,twocolumn,floatfix,10pt]{revtex4-2}
\usepackage[utf8]{inputenc}
\usepackage{amsmath,amssymb}
\usepackage{empheq}
\usepackage{lipsum}
\usepackage{mathtools,cuted}
\usepackage{esvect}
\usepackage{multirow}
\usepackage{xcolor}
\usepackage{soul}
\usepackage[export]{adjustbox}
\usepackage{graphicx}% Include figure files
\usepackage{dcolumn}% Align table columns on decimal point
\usepackage{bm}% bold math
\usepackage{physics}
\usepackage[mathlines]{lineno}% Enable numbering of text and display math
\usepackage[colorlinks=true,citecolor=blue]{hyperref}
\usepackage[capitalise]{cleveref}
\usepackage{bbm}
\usepackage{orcidlink}

\newcommand{\rr}[0]{\mathbf{r}}

\newcommand{\tenham}[0]{\hat{\mathcal{H}}}

\begin{document}
	
	\title{Tensor-network methodology for real-space super-moir\'e excitons}

    \author{Anouar Moustaj\,\orcidlink{0000-0002-9844-2987}}
    \thanks{These authors contributed equally}
     \affiliation{Department of Applied Physics, Aalto University, 02150 Espoo, Finland}
    \author{Yitao Sun\,\orcidlink{0009-0002-9479-7147}}
    \thanks{These authors contributed equally}
     \affiliation{Department of Applied Physics, Aalto University, 02150 Espoo, Finland}
    \author{Tiago V. C. Ant\~ao\,\orcidlink{0000-0002-7269-5513}}
    \affiliation{Department of Applied Physics, Aalto University, 02150 Espoo, Finland}
    \author{Lumen Eek\,\orcidlink{0009-0009-1233-4378}}
     \affiliation{Institute of Theoretical Physics, Utrecht University, Utrecht, 3584 CC, Netherlands}
    \author{Jose L. Lado\,\orcidlink{0000-0002-9916-1589}}
     \affiliation{Department of Applied Physics, Aalto University, 02150 Espoo, Finland}
 
	\date{\today}
	
	\begin{abstract}
    Computing excitonic spectra in quasicrystal and super-moir\'e systems constitutes a formidable challenge due to the exceptional size of the excitonic Hilbert space. Here, we demonstrate a tensor-network method for the real-space Bethe-Salpeter Hamiltonian, allowing us to access the spectra of an excitonic $10^{18}$-dimensional Hamiltonian, and enabling the direct computation of bound-exciton spectral functions for systems exceeding one billion lattice sites, 
    several orders of magnitude beyond the capabilities of conventional approaches. Our method combines a tensor-network encoding of the real-space Bethe-Salpeter Hamiltonian with a Chebyshev tensor network algorithm. This strategy bypasses explicit storage of the Hamiltonian while preserving full real-space resolution across widely different length scales.
    We demonstrate our methodology for one- and two-dimensional super-moir\'e systems, achieving the simultaneous resolution of atomistic and mesoscopic structures in the excitonic spectra in billion-size systems, showing
    exciton miniband formation and moir\'e-induced spatial confinement. 
    Our results establish a real-space methodology
    enabling the simulation of excitonic physics in large-scale 
    quasicrystal and super-moir\'e quantum matter.
	\end{abstract}
	
	\maketitle{}
		
\section{Introduction}

    Two-dimensional (2D) van der Waals materials provide an excellent platform for engineering excitonic matter owing to their enhanced Coulomb interactions.
    This has led to pronounced excitonic features that remain stable even at room temperature \cite{He2014TightlyWSe2,Ugeda2014GiantSemiconductor,Mueller2018ExcitonSemiconductors}, making these platforms ideal for technological applications \cite{Yin2012Single-LayerPhototransistors,Lopez-Sanchez2013UltrasensitiveMoS2,Pospischil2014Solar-energyDiode,Ross2014ElectricallyJunctions,Lee2014AtomicallyHeterointerfaces,Amani2015Near-unityMoS2}. 
    Importantly, control over twist angle and lattice mismatch in multilayer transition metal dichalcogenides (TMD) has enabled the realization of moir\'e excitons \cite{Yu2017MoireLattices,Seyler2019SignaturesHeterobilayers,Ruiz-Tijerina2020TheoryHeterobilayers,Regan2022EmergingHeterobilayers,delaTorre2025AdvancedExcitons,Westenberg2025Real-SpaceMoS2,Deng2025} by producing an effective potential landscape that modulates the coupling of electron and hole pairs across both layers, 
    % \TA{repeated excitons in this sentence. Perhaps... ``has allowed for a physical realization of moiré excitons [...] by producing an effective potential landscape that modulates the coupling of electron and hole pairs across both layers", or similar}
    leading to emergent phenomena, such as periodic confinement and excitonic minibands. These excitons can also become strongly localized at moir\'e trapping sites, forming programmable arrays of quantum emitters or allowing the engineering of exciton lattices that function as artificial quantum simulators \cite{Yu2017MoireLattices,He2015SingleSemiconductors,Deng2025}.

    Microscopically, excitonic phenomena can be predicted from the Bethe–Salpeter equation (BSE), which incorporates the single-particle electronic structure and the screened electron–hole interaction \cite{Frenkel1931OnI,Wannier1937TheCrystals,Keldysh1979CoulombII,Chernikov2014ExcitonWS2,Marino2018Quantum-electrodynamicalDichalcogenides,Ljungberg2015Cubic-scalingSystems,Blase2020TheChemistry,Deslippe2012BerkeleyGW:Nanostructures}. 
    Dedicated tight-binding BSE packages include Xatu~\cite{Uria-Alvarez2024EfficientCode}, WanTiBEXOS~\cite{Dias2023WanTiBEXOS:Solids}, and WannierBSE~\cite{Peng2019DistinctiveThermalization}. These construct the BSE in a restricted basis of selected valence and conduction bands, typically in momentum space for periodic systems, with a basis size scaling as $N^D_kN_vN_c$, where $N_k$ is the number of sampled momenta in the Brillouin zone, and $N_v$ and $N_c$ are the numbers of valence and conduction bands. Consequently, they do not retain the complete real-space electron--hole basis, limiting their applicability to arbitrary finite or strongly inhomogeneous systems. QNANO~\cite{Cygorek2020AtomisticDots} instead combines tight binding with configuration interaction, allowing correlations beyond the single electron--hole sector (e.g.\ charged complexes) at the expense of the combinatorial growth of the many-body Hilbert space.
    While for periodic systems a momentum space formulation of excitons is successful,
    for incommensurate modulations, such as those arising in moir\'e materials,
    a fully real-space formulation for excitonic effects becomes necessary.
    In these cases, conventional methods face severe computational limitations, as even at the single-particle level, real-space methods scale poorly with system size \cite{Wannier1937TheCrystals,Wu2018TheoryHeterobilayers,Herrera-Gonzalez2025MoireReview,Uria-Alvarez2024EfficientCode,Ljungberg2015Cubic-scalingSystems,Blase2020TheChemistry}.
    More specifically, for a single-particle Hamiltonian defined on $N$ real-space sites, a dense exact diagonalization scales as $\mathcal{O}(N^3)$ in time and $\mathcal{O}(N^2)$ in memory for each relevant band sector \cite{Golub2013SymmetricProblems}, while sparse or polynomial-expansion methods can achieve more favorable scaling \cite{Golub2013LargeProblems,Weie2006TheMethod}, where an explicit sparse-matrix representation has at best memory scaling that is linear in system size \cite{Saad2003SparseMatrices}.
    A number of efficient real-space single-particle methodologies exploit this
    sparse-matrix scaling, including for spectral properties and quantum transport~\cite{Joao2020KITE:Heterostructures,Groth2014Kwant:Transport,Li2022TBPLaS:Simulation,Papior2017Sisl:V0.9.2,Dean2020Pybinding,Uria-Alvarez2024Tightbinder:Solids}. 
    For excitons, the situation is more demanding, since the two-particle Hilbert space scales quadratically with the number of lattice sites. Therefore, a direct dense treatment of the BSE would scale as $\mathcal{O}(N^6)$ in time and $\mathcal{O}(N^4)$ in memory, while sparse methods can at best yield an $\mathcal{O}(N^2)$ memory scaling. In super-moir\'e materials, where characteristic length scales can exceed those of standard moir\'e systems by orders of magnitude, the effective lattice may involve millions or even billions of sites \cite{Chen2019EvidenceSuperlattice,Zhu2020ModelingHeterostructures}, rendering brute-force approaches computationally prohibitive.
     
    In the single-particle case, quantum many-body solvers based on tensor networks (TNs) have enabled the treatment of exponentially large Hamiltonians \cite{Fumega2025CorrelatedAlgorithm,Sun2025Self-consistentSites,Antao2025TensorMosaics,Moustaj2026TensorSystems,Sun2026Real-spaceNetworks,Antao2026TensorApplications} with a much cheaper $\mathcal{O}(\log N)$ memory scaling. These approaches avoid explicit storage of large matrices by representing operators and states with TNs: many-body factorizations that permit algebraic manipulations in exponentially large Hilbert spaces
    \cite{PhysRevLett.69.2863,RevModPhys.93.045003,Bauls2023,Ors2014,Schollwck2011,Stoudenmire2012,Fishman2022TheCalculations,Fishman2022CodebaseITensor,PhysRevX.10.041038,Huggins2019,Ors2019,Chan2016}. The required tensor representations are constructed using tensor cross interpolation techniques \cite{Oseledets2010,Oseledets2011,Ritter2024QuanticsFunctions,PhysRevB.110.035124,NunezFernandez2025LearningLibraries,QuanticsTCI.jl,TensorCrossInterpolation.jl}.
    Similarly, the employment of TNs to encode classical functions has led to speedups of several orders of magnitude across diverse computational settings \cite{2026arXiv260103035W,PhysRevX.13.021015,PhysRevB.107.245135,PhysRevX.12.041018,10.21468/SciPostPhys.18.1.007,PhysRevB.110.035124,Jeannin2025,Rohshap2025Two-particleEquations}, including applications in computational chemistry \cite{Jolly2025} and dynamical simulations \cite{Peddinti2024,Gourianov2025,Niedermeier2026SolvingRepresentation,boucomasGP,chenGP,2025arXiv250701149C}. These developments have enabled, among others, the study of quasicrystalline mosaics of topological Chern states \cite{Antao2025TensorMosaics}, the self-consistent solution of interacting super-moir\'e systems at scales exceeding one billion lattice sites \cite{Sun2025Self-consistentSites}, and the direct computation of spatially-dependent momentum-resolved spectral function of large nonperiodic systems \cite{Moustaj2026TensorSystems}.

    Here, we demonstrate a tensor-network (TN) based methodology to access the real-space exciton spectra in super-moire systems. Instead of targeting truncated excitonic bases, our method compresses and extracts spectral information from the complete real-space two-particle BSE Hamiltonian.
    Our methodology relies on combining a real-space TN 
    representation of the electron–hole Hilbert space, and tensor-network kernel polynomial 
    Chebyshev algorithms with a high-resolution spectral kernel.
    Our approach enables the solution of
    the real-space BSE on lattices exceeding one billion sites, both in 1D and 2D, surpassing by
    many orders of magnitude the capabilities of conventional BSE solvers \cite{Ljungberg2015Cubic-scalingSystems,Blase2020TheChemistry,Uria-Alvarez2024EfficientCode,Dias2023WanTiBEXOS:Solids,Peng2019DistinctiveThermalization}. We show that these techniques allow probing the emergence of exciton moire minibands, together
    with imaging the microscopic exciton localization, both in quasicrystalline and super-moire systems.

\section{Tensor Network Methodology}

    For concreteness, we formulate our algorithm in a minimal two-band model that captures excitonic physics, and we solve it with our TN methodology. We take an interacting Wannier Hamiltonian for the valence and conduction band of a semiconductor as $\hat{H} = \hat{H}_c + \hat{H}_v + \hat{H}_U$, with
     \begin{equation}\label{Eq: MB Hamiltonian}
         \begin{split}
         \hat{H}_{b} &= \sum_{l,m} t^{}_{b,lm} c_{b,l}^\dagger c_{b,m}^{} +  \sum_{l} V_{b,l}^{} c_{b,l}^\dagger c_{b,l}^{}, \\
         \hat{H}_U &= \sum_{l} U^{}_l c_{c,l}^\dagger c_{c,l}^{} c_{v,l}^\dagger c_{v,l}^{},
         \end{split}
     \end{equation}
    where $c^\dagger_{c,l}$ ($c^\dagger_{v,l}$) creates an electron in the Wannier conduction (valence) band at site $l$. 
    We consider spinless electrons in this work for simplicity.
    The band index is denoted by $b=c,v$ for conduction and valence, respectively. The hopping amplitudes are given by $t_{lm}$ 
    % \YT{I notice here $t^{}_{b,lm}$ seems to be a constant for electrons and holes, better include $b$ also in the hopping to distinguish}
    , while $V_{c,l}=-V_{v,l}$ represents the on-site potential, and $U_l$ describes the local screened electrostatic interaction between a conduction-band Wannier state and a valence-band Wannier state. This Hamiltonian 
     stems from projecting a microscopic interacting electron model onto the low-energy subspace spanned by the top valence and bottom conduction bands, performing a Wannier transformation to real space, and retaining the local Coulomb interaction between valence and
     conduction bands. 
%     Without loss of generality, we omit exchange terms, as they vanish for the chosen local interaction. 
    \begin{figure}[!hbt]
        \centering
        \includegraphics[width=\columnwidth]{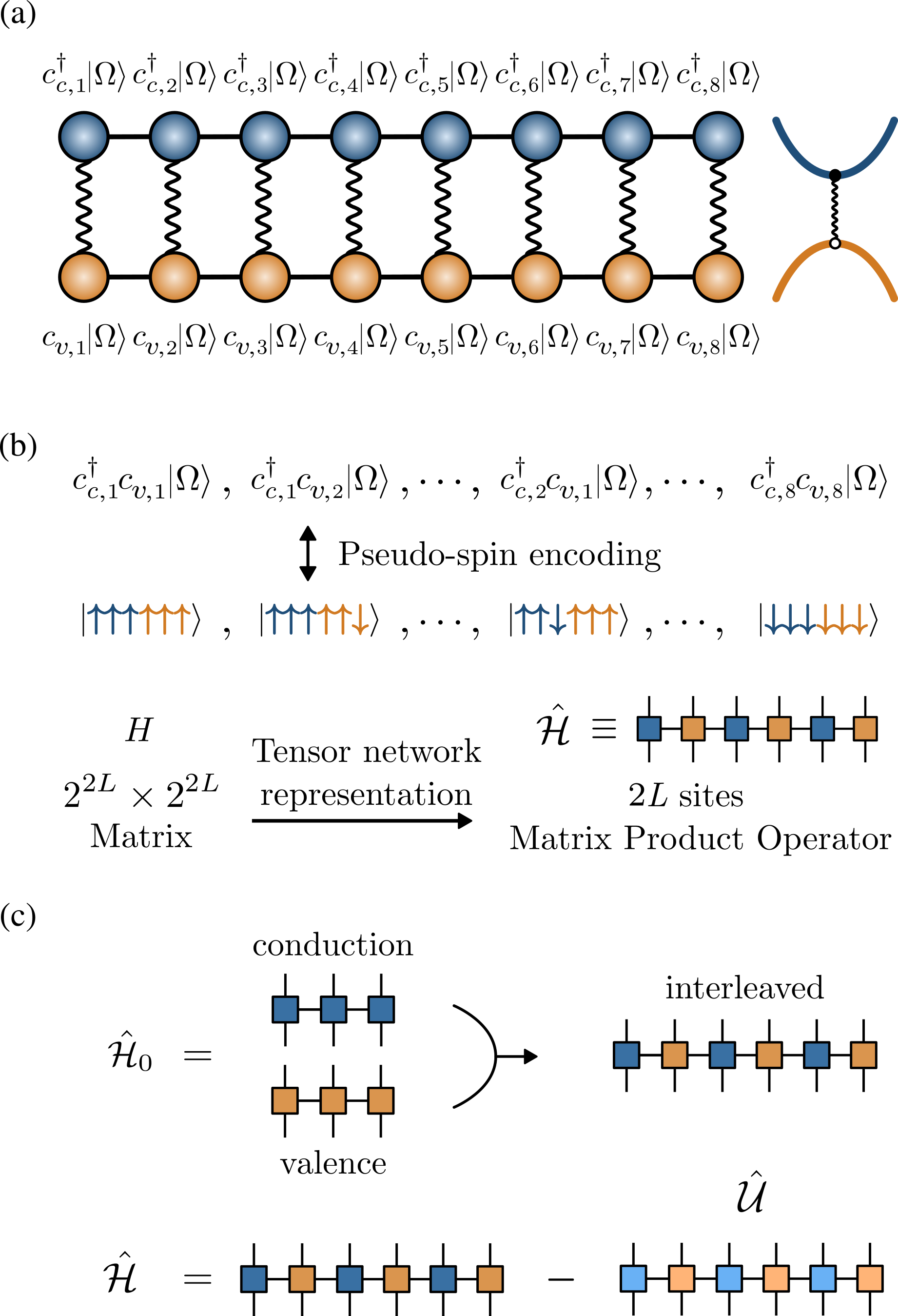}
        \caption{(a) Schematic of the exciton Hamiltonian, with purple corresponding to the conduction band and light brown to the valence band, together with their corresponding single-particle basis. (b) The encoding of the two-particle $2^{2L}\times2^{2L}$ Hamiltonian matrix into a $2L$ pseudo-spin MPO. (c) The combination of the single-particle Hamiltonians and the interaction kernel using interleaved ordering. 
        % \LE{We now use both $c_v$ and $v$ as valence band operators (similar for conduction). We should pick one convention and stick with it}
        } 
        \label{fig: Hamiltonian MPO}
    \end{figure}
     Our objective is to solve the effective two-particle Hamiltonian obtained from the BSE by projecting the many-body Hamiltonian \cref{Eq: MB Hamiltonian} onto the variational exciton state $\ket{l,m} = c^\dagger_{c,l} c_{v,m} ^{}\ket{\Omega}$, where $\ket{\Omega} = \prod_{l=1}^N c_{vl}^\dagger \ket{0}$ is the Fermi sea, filled with $N$ electrons. 
     The specific matrix takes the form
     \begin{equation}\label{Eq: Exciton Matrix}
         \hat{H}_X =\hat{T_c}\otimes\mathbbm{1} - \mathbbm{1}\otimes\hat{T_v} - \hat{U},
     \end{equation}
     where $\hat{T}_c$ contains both the conduction hopping term and an additional contribution from the local Coulomb interaction, $\hat{T}_v$ is the contribution arising from the valence band, and $\hat{U}$ couples electrons and holes on the same site. Its matrix elements are $U_{klmn}=U_k\delta_{kl}\delta_{mn}\delta_{ln}$ and, as such, cannot be written as a product operator (see App.~\ref{App: exciton Hamiltonian} for details).

     The key strategy of our methodology lies in the encoding of $2^L$ real-space sites of the Wannier tight-binding Hamiltonian into $L$ pseudo-spin indices \cite{Fumega2025CorrelatedAlgorithm,Sun2025Self-consistentSites,Antao2025TensorMosaics,Moustaj2026TensorSystems} for each sector of the BSE Hamiltonian, effectively resulting in a system of $2L$ pseudo-spin degrees of freedom. In this pseudo-spin basis, the single-particle Hamiltonian of each sector can be written as a TN matrix product operator(MPO) of the form $\mathcal{H}_{(s_1,s_2,...,s_L),(s'_1,s'_2,...,s'_L)} = {\Gamma}^{s_1,s_1'}_1{\Gamma}^{s_2,s_2'}_2\cdots {\Gamma}^{s_L,s_L'}_L$.
    Each tensor $\Gamma_r$ carries four indices: two virtual indices, which are contracted with neighboring tensors, and two physical indices associated with the local 2D Hilbert space $s_r,s'_r$.
    The virtual indices have dimension $\chi_r$, referred to as the bond dimension, which determines the expressiveness of the local tensor.
    The explicit form of the tensors can be obtained with quantics tensor cross interpolation \cite{Ritter2024QuanticsFunctions,2026arXiv260103035W,PhysRevX.13.021015,PhysRevB.107.245135,PhysRevX.12.041018,10.21468/SciPostPhys.18.1.007,PhysRevB.110.035124,Jeannin2025}, 
    enabling the representation of spatially varying Hamiltonians as a TN. 
    Considering the first two terms of \cref{Eq: Exciton Matrix}, a straightforward many-body encoding consists of two independent pseudo-spin chains of length $L$, which together form a composite chain of length $2L$, reflecting the direct-product structure of the corresponding operators. 
    However, in this pseudo-spin representation, the interaction operator is expressed as 
    $\hat{\mathcal{U}} = \prod_{r=1}^L \left( \frac{1}{2} \mathbbm{1}_r \otimes \mathbbm{1}_{r+L} + 2  \sigma^{z}_r   \sigma^z_{r+L} \right)$.
    This specific form couples the two pseudo-spin chains by sequentially ordering all electron sites before all hole sites. 
    Such a layout leads to an exponential growth of the bond dimension at the interface between the two MPOs as the system size increases. 
    To mitigate this, we use a more efficient interleaved ordering \cite{NunezFernandez2025LearningLibraries,2026arXiv260103035W} of the sites, which maintains a constant bond dimension for the interaction operator regardless of the system size. In essence, rather than linking the individual pseudospin blocks with a single internal bond at their interface, we adopt an interleaved ordering in which the tensor cores are rearranged so that each local electron pseudospin degree of freedom is directly coupled to its corresponding hole pseudospin degree of freedom.
    The MPO construction of the Hamiltonian is illustrated in \cref{fig: Hamiltonian MPO}. The technical details of the implementation of the interleaved BSE Hamiltonian is discussed in App.~\ref{App: interleaved order}.

    % \YT{We can also mention it is not really exciton LDOS, but a real-space measure of the bound-exciton spectral weight, as in the response}
    The excitonic spectral signatures can be obtained
    with a Chebyshev TN kernel polynomial algorithm (KPM) \cite{Weie2006TheMethod}. 
    This allows for a direct calculation of a real-space measure of the bound-exciton spectral weight, or local density of states (LDOS), $\rho(\rr,\omega) = \bra{\rr,\rr}\delta(\omega-\tenham)\ket{\rr,\rr}$. 
    The exciton LDOS is expanded as
    \begin{equation*}
    \rho(\rr, \Tilde{\omega}) = \frac{1}{\pi\sqrt{1-\Tilde{\omega}^2}}\left[1 + 2\sum_{n=1}^\infty\mu_n(\rr)T_n(\Tilde{\omega})\right],
    \end{equation*}
    where $T_n(\Tilde{\omega})$ are Chebyshev polynomials of a rescaled energy $\Tilde{\omega}\in(-1,1)$ and $\mu_n(\rr)$ are the Chebyshev moments. 
    These constitute the core building blocks of the present methodology and are constructed according to the recursion relations \cite{Weie2006TheMethod}
    \begin{equation}\label{Eq: Chebyshev moments}
    \begin{split}
        \mu_n(\rr) &=\bra{\rr,\rr}\ket{\nu_n}, \\
        \ket{\nu_n} &= 2\tenham\ket{\nu_{n-1}} - \ket{\nu_{n-2}}, \\
        \ket{\nu_0} &= \ket{\rr,\rr}, \ \ \ \ket{\nu_1}=\tenham\ket{\nu_0},
    \end{split} 
    \end{equation}
    where $\tenham$ is a rescaled Hamiltonian whose eigenvalues are in $(-1,1)$. In practice, the expansion is truncated to $N_\mu$ terms. Within the KPM framework, the inclusion of the Jackson kernel \cite{Jackson1912} yields an effective energy resolution of order $1/N_\mu$, thereby setting the scale of the associated spectral broadening \cite{Weie2006TheMethod}.
    For 2D systems, however, the rapid growth of the bond dimension with the number of Chebyshev moments necessitates an alternative strategy. To keep $N_\mu$ moderate while maintaining spectral resolution, we employ a high-order delta-Chebyshev (HODC) kernel \cite{Yi2025ADensitiesb}, which enables independent control of the KPM truncation error through additional parameters $\eta$ and $m$, while achieving a superior pointwise convergence. 
    % $\eta^m\sim1/N_\mu^m$(see SM \cite{SuppMat} for details).
    The construction of the HODC kernel proceeds as follows: the $m^\text{th}$ order rational approximation of $\delta(\omega-E)$ is obtained from 
    \begin{equation}\label{Eq: rational approximant of delta}
        K_\eta(\omega,E) = -\frac{1}{\pi}\sum_{l=1}^m\Im\frac{w_l}{\omega-E+\eta z_l}.
    \end{equation}
    where the weights $w_l$ and poles $z_l$ are both complex numbers and need to satisfy specific matching conditions for this approximation to hold. This is enforced by solving the Vandermonde system of linear equations \cite{Colbrook2021ComputingOperators}
    \begin{equation}\label{Eq: Vandermonde system}
        \begin{bmatrix}
        1 & 1 & \dots & 1 \\
        z_1 & z_2 & \dots & z_m \\
        \vdots & & & \vdots \\
        z_1^{m-1} & z_2^{m-1} & \dots & z_m^{m-1}
    \end{bmatrix}
    \begin{bmatrix}
        w_1 \\
        w_2 \\
        \vdots \\
        w_m
    \end{bmatrix}
    =
    \begin{bmatrix}
        1 \\
        0 \\
        \vdots \\
        0
    \end{bmatrix}.
    \end{equation}
    The solutions $w_l$ result in a kernel $K_\eta(\omega ,E)$ that approximates $\delta(\omega-E)$ with an accuracy of order $\mathcal{O}(\eta^m)$ \cite{Yi2025ADensitiesb}.
    In practice, for $m<8$, the solutions $\omega_l$ can be calculated to double precision. 
    Then, to approximate the kernel $K_\eta(\omega,E)$, we can use a Chebyshev expansion  
    \begin{equation}
        K_\eta(\omega,E)  \approx \sum_{k=0}^{N_\nu - 1}\nu_k(\omega,\eta)T_k(E),
    \end{equation}
    where the number of moments $N_\nu$ should be of order $\mathcal{O}(1/\eta)$ to achieve a precise approximation. 
    With this, we can calculate the LDOS at site $i$ of a Hamiltonian $H$ as
    \begin{equation}
    \begin{aligned}
        D_i(\omega) = \bra{i}\delta(\omega - H)\ket{i}
        \approx \sum_{k=0}^{N_\nu-1} \nu_{k}(\omega, \eta)\bra{i}T_{k}\left(H\right)\ket{i},
    \end{aligned}
    \label{eq:ldos}
    \end{equation}
    where $T_k(H)$ is the $k-$th order Chebyshev polynomial of $H$. Finally, to acquire the expansion coefficients $\nu_k(\omega, \eta)$ for given $\omega$ and $\eta$, we first solve Eq.~\ref{Eq: Vandermonde system} to obtain the weights $w_l$. Subsequently, we sample $n$ values of $E$ at the Chebyshev nodes (roots of the $n$-th Chebyshev polynomial),
    \begin{equation}
    E_j = \cos\left( \frac{(j + \frac{1}{2})\pi}{n} \right), \quad j = 0, 1, \dots, n-1.
    \end{equation}
    The coefficients $\nu_k$ are then obtained by projecting the function onto the Chebyshev basis. This projection is mathematically equivalent to a Discrete Cosine Transform (DCT):
    \begin{equation}
    \nu_k(\omega, \eta) = \frac{2 - \delta_{k0}}{n} \sum_{j=0}^{n-1} K_{\eta}(\omega,E_j)\cos\left( \frac{k(j+\frac{1}{2})\pi}{n} \right),
    \end{equation}
    %By utilizing the fast cosine transform, this summation can be performed with a computational cost of $\mathcal{O}(p \log p)$~\cite{Trefethen2019}\JL{I guess that this would still be doable with brute force cosine transform?}. 
    This allows for the rapid acquisition of all $N_\nu$ terms of $\nu_k(\omega, \eta)$, which are subsequently used to evaluate the exciton LDOS via Eq.~\ref{eq:ldos}. 

    The motivation behind the HODC is the superior convergence performance. Here, the Dirac delta function is replaced by an $m$th-order regularized kernel with energy broadening $\eta$. 
    For sufficiently smooth spectral densities and away from singular points, this regularization yields a pointwise error of order $O(\eta^m)$ in the regularized LDOS. 
    Since the Chebyshev expansion of the regularized kernel requires a polynomial order that scales approximately as $N_\mu\sim O(1/\eta)$ for fixed accuracy, this yields an effective convergence rate of $O(N_\mu^{-m})$ with respect to the Chebyshev order. 
    This should be contrasted with the standard Jackson-kernel KPM, whose pointwise convergence at smooth points is typically second order, $O(N_\mu^{-2})$. 
    
    Despite its superior convergence properties, the primary drawback of the HODC kernel is the lack of guaranteed positivity, in contrast to traditional truncation kernels (e.g., Jackson or Lorentz kernels)~\cite{Weie2006TheMethod,Jackson1912}. This characteristic may introduce minor numerical artifacts in the calculation of the excitonic LDOS, which is physically required to be non-negative. However, these errors are typically sufficiently small to be neglected when weighed against the kernel's rapid convergence, as shown in \cref{Sec: Num Bench}.
    
    We now demonstrate the capabilities of our TN framework by applying it to two representative examples that require the simultaneous resolution of both macroscopic and atomistic length scales.

\section{Results}

\subsection{1D incommensurate super-moir\'e potential}

\begin{figure*}[!t]
        \centering
        \includegraphics[]{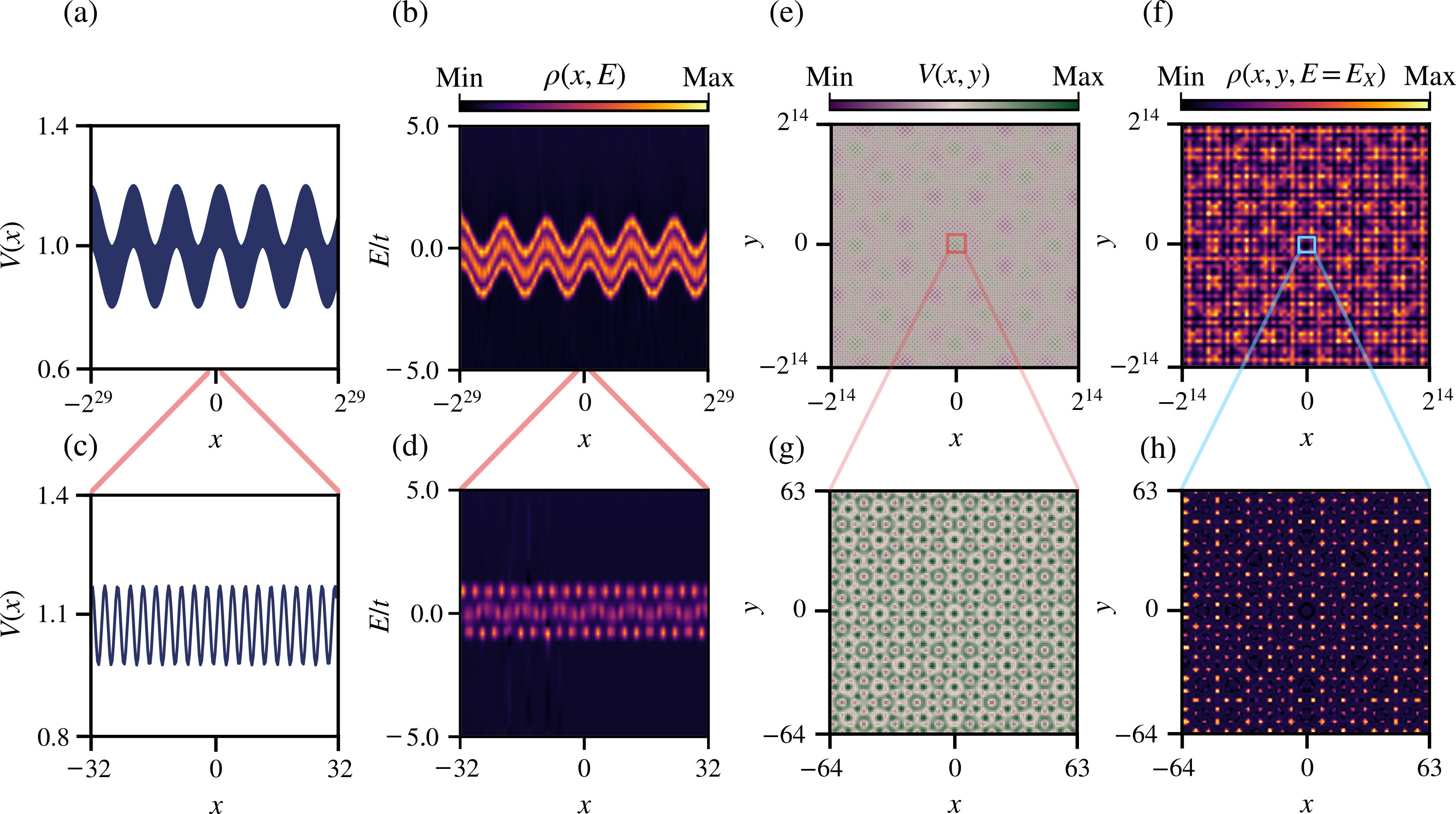}
        \caption{Results of large-scale calculations for both the 1D and 2D examples, respectively. (a,c) 1D Super-moir\'e on-site potential modulation, with a corresponding zoom of the central region of length $\ell = 64$. The outlined region is not shown to scale with respect to the full domain and serves only as a visual guide. (b,d) Spectral band in real space of the bound super-moir\'e excitons, together with a zoom of the central region of linear extent $\ell = 64$. The parameters used for simulation are $t = -1$, $V_0 = 3.0$, $U_0 = 6.0$. The system size is $N=2^{30}$, and the TN parameters are $\chi_{\rm max}=200$ and $\rm tol = 10^{-10}$. (e,g) 2D Super-moir\'e on-site potential, with a zoom of the central region of area $A=128\times128$. The square outline is not drawn to scale relative to the full domain and is included as a visual guide. (f,h) Corresponding 2D exciton LDOS $\rho(x,y,\omega=E_X)$, evaluated at the bound-exciton energy $E_X=-1$, together with a zoom of the central region of area $A=128\times128$. The parameters used for simulation are $t = -0.5$, $V_0 = 6.0$, $U_0 = 10$. 
        % \YT{The paras for 2D excitons are not consistent with in the text, double check it}
        The system size is $N_x=N_y=2^{15}$ and the TN parameters are $\chi_{\rm max}=400$ and $\rm tol = 10^{-11}$. In all LDOS panels, color encodes only the normalized local spectral weight $\rho$ at the plotted energy, from dark (small LDOS) to bright (large LDOS); intermediate colors indicate intermediate intensity and do not denote a distinct phase, state, or additional physical quantity.}
        \label{fig: 1D 2D super moire}
    \end{figure*}

    The first example is a 1D system subject to a modulated single-particle on-site potential of the form
    \begin{equation}\label{Eq: super-moire 1D modulation}
    \begin{split}
        V_i &= V_0\left[1 + 0.1\cos\left(\frac{2\pi x_i}{b_1}\right) + 0.1\cos\left(\frac{2\pi x_i}{b_2}\right)\right],
    \end{split}
    \end{equation}
    where the two modulations $b_1=3\sqrt5/2$ and  $b_2=\sqrt3 N/10$ are incommensurate with each other. 
    This model captures moir\'e physics in quasi-one-dimensional van der Waals materials \cite{Island2017,Ma2022MultipleTaSe3,Balandin2022One-dimensionalField} or
    multi-walled nanotubes \cite{Birkmeier2022,Nishida2025,Tomio2012InterwallNanotubes,Gong2017HighPerformanceNanohybrids,aha2025}.
    The first cosine induces a small-scale modulation, while the second induces a large-scale one. We also consider nearest-neighbor uniform hopping $t=-1$, a constant electron-hole interaction strength $U=6$, and an on-site potential strength $V_0=3$ (in units of $|t|$). Figure \ref{fig: 1D 2D super moire}(a) shows the large-scale modulation \cref{Eq: super-moire 1D modulation} for a system of $N=2^{30}$ sites and \cref{fig: 1D 2D super moire}(b) the corresponding large-scale LDOS as a function of energy. The isolated exciton band in real-space closely follows the spatial profile of the on-site modulation potential. In \cref{fig: 1D 2D super moire}(c) and \cref{fig: 1D 2D super moire}(d) provide a zoom into the central region of length $\ell=64$ of the same quantities as in \cref{fig: 1D 2D super moire}(a,b). The spatial modulation and its imprint in the LDOS are also well resolved at the atomic scale.
    Furthermore, the incommensurability of the small-scale modulation frequency gives rise to miniband splitting, which is visible at both length scales through the modulated small gap.
    The spatial modulation changes the local electron-hole environment and shifts the local excitonic energy. 
    When excitonic states associated with different local environments hybridize, the LDOS of the Frenkel exciton develops miniband-like splittings, a feature clearly visible in \cref{fig: 1D 2D super moire} (b,d). 
    This effect is not tied to a single supercell periodicity, but appears as a hierarchy of features visible on both the long modulation scale and the shorter local scale.

    Finally, we note that for the 1D incommensurate chain, the spectral half-width is $a \simeq 5.6$. Thus, using the Jackson kernel with $N_\mu=150$, we reach an effective energy broadening of $\Delta E \simeq \pi a / N_\mu \simeq 0.12$. The TN parameters are $\chi_{\rm max}=200$ and $\rm tol = 10^{-10}$.

\subsection{2D quasiperiodic incommensurate super-moir\'e potential}

    We now demonstrate the methodology on a 2D super-moir\'e quasicrystal. 
    We consider a square lattice with an on-site potential exhibiting an eight-fold quasiperiodic, incommensurate modulation at both small and large scales. 
    \begin{equation}\label{Eq: V potential 8fold}
    V(\mathbf{r}) = V_0\left[1 + \sum_{n=1}^4\left(\Delta_\alpha \cos\left(\alpha \mathbf{k}_n\cdot \rr \right) + \Delta_\beta \cos\left(\beta \mathbf{k}_n\cdot \rr\right)\right)\right],
    \end{equation}
    where $\Delta_\alpha$ sets the modulation amplitude at the atomic scale, while $\Delta_\beta$ determines the super-moir\'e modulation amplitude. The eightfold pattern is created by the superposition of four wave vectors given by $\mathbf{k_n}=R^n(\pi/4)\left[2\pi, 0\right]^T$, with $R(\pi/4)$ denoting the 2D rotation matrix corresponding to an angle of $\pi/4$ radians. This potential can result from the superposition of two square lattice monolayers rotated at $45^\circ$ relative to each other, as in twisted GeX/SnX moir\'e superlattices \cite{Xu2025EngineeringSuperlattices}. 
    We set $\Delta_\alpha = 0.15$, $\Delta_\beta = 0.05$, and take moir\'e length scales $\alpha = 2/(5\sqrt{5})$ and $\beta = 16/(\sqrt{3}N_x)$. The uniform nearest-neighbor hopping is $t=-0.5$, the on-site modulation potential amplitude $V_0=6$, and the local Coulomb interaction $U_0=10$. The large-scale modulation landscape of the on-site potential is shown in \cref{fig: 1D 2D super moire}(e) for a system of size $N_x=N_y=2^{15}$, corresponding to more than a billion lattice sites. In \cref{fig: 1D 2D super moire}(f), the corresponding LDOS at the bound exciton energy $E_X=-1$ is depicted, where one can observe that the spectral weight peaks near the minima of the on-site potential, indicating spatial confinement of the excitons in a pattern with eight-fold rotational symmetry. In \cref{fig: 1D 2D super moire}(g) and \cref{fig: 1D 2D super moire}(h), 
    a zoom into a central region of area $A=128\times128$ is shown for the on-site potential and LDOS respectively. Similarly to the super-moir\'e length scale, the same eightfold-symmetric confinement pattern at the minima of the potential can be observed at the atomic length scale. 
    
\begin{figure*}[!t]
        \centering
        \includegraphics[]{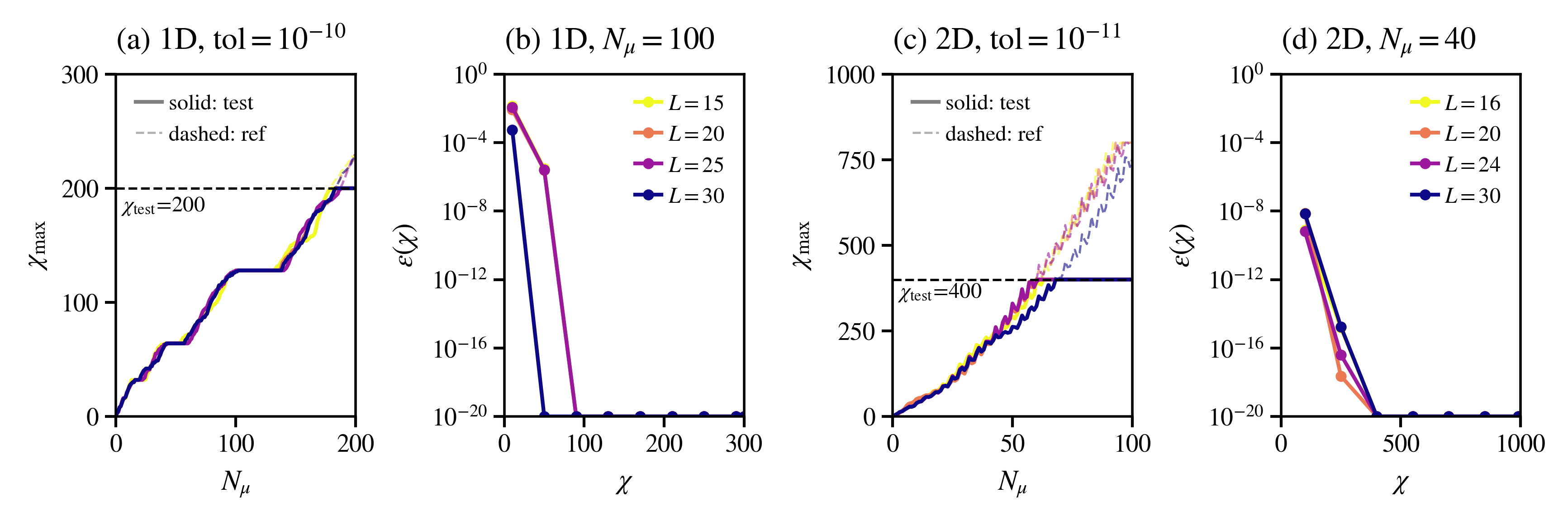}
        \caption{Numerical benchmarking of the exciton LDOS (main-text summary; the complete benchmark set is given in the App.~\ref{App: extra benchmarks}, \cref{fig: full benchmark}). Panels (a,c) show the growth of the maximum bond dimension $\chi_\mathrm{max}$ with Chebyshev order $N_\mu$ for the test (solid) and reference (dashed) recursions; the horizontal dashed line marks the test cap $\chi_\mathrm{test}$. Panels (b,d) show the normalized RMS LDOS error $\varepsilon(\chi)$ [\cref{Eq: RMS error}] between the LDOS computed at bond dimension $\chi$ and a converged high-$\chi$ reference, at a fixed number of Chebyshev moments $N_\mu$. The left pair corresponds to the 1D incommensurate chain [(a) $\chi_\mathrm{test}=200$, $\chi_\mathrm{ref}=500$, $\mathrm{tol}=10^{-10}$; (b) $N_\mu=100$] and the right pair to the 2D eight-fold quasiperiodic lattice [(c) $\chi_\mathrm{test}=400$, $\chi_\mathrm{ref}=800$, $\mathrm{tol}=10^{-11}$; (d) $N_\mu=40$]. Colors run from yellow to blue with increasing system size $N=2^L$: in 1D, $L\in{15,20,25,30}$ in (a) and ${15,20,25,30}$ in (b); in 2D, $L=2L_x$ (square lattices) with $L_x\in{8,10,12,15}$ in (c,d). In all panels the exciton is probed at the deepest on-site potential well, $X=\arg\min_x V(x)$. The 1D recursion converges at a modest bond dimension, whereas the 2D case requires substantially larger $\chi$ and saturates its test cap already at low $N_\mu$.
        % \YT{don't forget to unify $N_\mu$ also in the main text}
        }
        \label{fig: convergence mps Ncheb}
    \end{figure*}

    % \JL{what are the numbers in parentheses and the others?}
    For the 2D model, the spectral half-width is $a \simeq 14.4$.
    As shown in \cref{fig: convergence mps Ncheb}(c), the bond dimension $\chi$ saturates at $N_\mu \simeq 50$--$60$ for $\rm tol = 10^{-11}$, making the calculation bond-dimension limited and precluding the use of a large number of moments. By using the HODC kernel with $m = 3$, $N_\mu = 40$ and $\eta = 0.03$, we reach physical energy broadening of $\Delta E = a\eta \simeq 0.43$.
    For comparison, the Jackson kernel at the same $N_\mu = 40$ would give $\Delta E \simeq \pi a/N_\mu
    \simeq 1.13$, i.e.\ more than twice broader; the HODC kernel is thus the
    appropriate choice precisely in the bond-dimension-limited regime where $N_\mu$ cannot be made large. The TN parameters used are $\chi_{\rm max}=400$ and $\rm tol = 10^{-11}$.\\

    Even though the model is idealized, it illustrates how incommensurate spatial structure can produce nontrivial real-space excitonic phenomenology. 
    The main physical effect captured in both the 1D and 2D systems is the redistribution and splitting of excitonic spectral weight by the incommensurate modulation.
    The spatial modulation changes the local electron-hole environment and shifts the local excitonic energy. In 2D, the LDOS reveals excitonic confinement in an aperiodic potential landscape across widely separated length scales. The spectral weight follows the underlying potential modulation at both the super-moir\'e and atomic scales, demonstrating the confinement of bound excitons at the potential minima. Such confinement is expected to arise in systems with aperiodically stacked van der Waals heterostructures.

\section{Numerical Benchmarking}\label{Sec: Num Bench}

    We now present numerical benchmarks to assess the performance and limitations of the current methodology. The first test will probe the convergence and accuracy of tensor-network MPS propagation, the second test probes the accuracy of the LDOS as a function of compression, and the third test will examine the behavior of the HODC kernel.

    \subsection{Tensor Network MPS propagation convergence}

    We first characterize the cost of the KPM propagation itself through the growth of the MPS bond dimension with Chebyshev order. Panels (a,c) of \cref{fig: convergence mps Ncheb} show the maximum bond dimension $\chi_\mathrm{max}$ reached by the Chebyshev recursion as a function of the order $N_\mu$, for a test recursion (solid, capped at $\chi_\mathrm{test}$) and a higher-tolerance reference (dashed, capped at $\chi_\mathrm{ref}$), with the exciton probed at the deepest on-site well $X=\arg\min_x V(x)$. In 1D [(a), $\chi_\mathrm{test}=200$, $\chi_\mathrm{ref}=500$, $\mathrm{tol}=10^{-10}$] the bond dimension grows steadily and the test recursion reaches its cap only near $N_\mu\approx180$, the reference retaining ample headroom below $\chi_\mathrm{ref}$; the 1D calculation is therefore not bond-dimension limited over the range of interest. In 2D [(c), $\chi_\mathrm{test}=400$, $\chi_\mathrm{ref}=800$, $\mathrm{tol}=10^{-11}$] the requirements are far more severe: $\chi_\mathrm{max}$ rises rapidly, the test recursion saturates its cap by $N_\mu\approx65$, and even the reference approaches $\chi_\mathrm{ref}$ by $N_\mu=100$, so the accessible number of Chebyshev moments is intrinsically limited, which is why we adopt the HODC kernel, resolving sharper spectral features at a low moment count. System sizes are $N=2^L$, with $L\in{15,20,25,30}$ in 1D and $L=2L_x$, $L_x\in{8,10,12,15}$ (square lattices) in 2D.
    
    While the bond-dimension growth reflects the complexity accumulated during the KPM propagation, it is more instructive to quantify the error directly in the resulting LDOS. This is done by calculating the relative RMS error
    \begin{equation}\label{Eq: RMS error}
    \epsilon(\chi)= \sqrt{\frac{\sum_{ij}|\rho_{ij}(\chi)-\rho_{ij}(\chi_\text{ref})|^2}{\sum_{ij}|\rho_{ij}(\chi_\text{ref})|^2}},
    \end{equation}
    where $\rho_{ij}$ abbreviates $\rho(x_i,\omega_j)$. Panels (b,d) show $\epsilon(\chi)$ between the LDOS computed at MPS bond dimension $\chi$ and a converged high-$\chi$ reference, at a fixed number of Chebyshev moments $N_\mu$ ($N_\mu=100$ in 1D, $N_\mu=40$ in 2D), for increasing system size $L$. In 1D (b) the error collapses to the numerical floor already at a modest bond dimension ($\chi\lesssim90$), growing only mildly with $L$. In 2D the decay is far slower and the error reaches the floor only near $\chi\sim 400$. The complementary dependence on the moment count $N_\mu$ and on the compression tolerance, together with the Chebyshev-vector infidelity, is presented in App.~\ref{App: extra benchmarks}.

    \begin{figure*}[!t]
        \centering
        \includegraphics{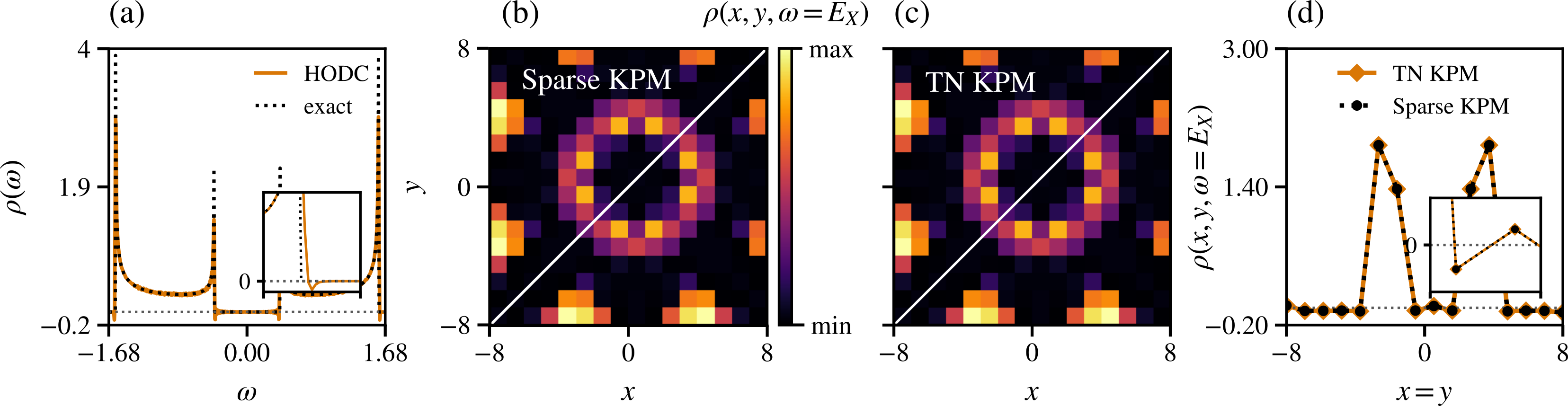}
        \caption{Illustration of HODC-kernel negativity and validation of the spatial features of the 2D exciton LDOS. Panel (a) shows the SSH-chain density of states $\rho(\omega)$ computed exactly and with sparse Chebyshev KPM using the HODC kernel; the inset highlights the small negative HODC oscillations near a van Hove singularity. Panels (b) and (c) show $\rho(x,y,\omega=E_X)$ for the 2D exciton problem computed with sparse-matrix KPM and tensor-network KPM, respectively, using the HODC kernel. The white diagonal marks the line $x=y$. Panel (d) compares the corresponding diagonal profiles $\rho(x,x,\omega=E_X)$, with the inset emphasizing the near-zero region. The 2D calculations use the same physical parameters as in the main calculations, but at $L_x=L_y=4$.}
        \label{fig: ldos hodc check}
    \end{figure*}

\subsection{The HODC kernel}

    The negative values of the HODC are a numerical artifact of the regularized spectral approximation, rather than a sign of negative physical spectral weight. 
    To demonstrate this explicitly, we first show in \cref{fig: ldos hodc check}(a) how these negative values appear even for a simple 1D SSH \cite{Su1979SolitonsPolyacetylene} chain near van Hove singularities. The DOS was computed for a system of $N=20000$ sites using stochastic trace KPM with sparse-matrix vector propagation, employing $N_\mu=500$, $\eta=0.003$, and $m=4$. Additionally, we compute the 2D exciton LDOS at a bound-exciton energy $E_X$ on a small system ($L_x = L_y = 4$) using both a sparse matrix algorithm and our tensor network algorithm.
    The result is shown in \cref{fig: ldos hodc check}. Panels (b) and (c) display the full spatial LDOS map
    $\rho(x,y,\omega=E_X)$ from sparse and tensor-network KPM, respectively, while panel (d) shows the diagonal cut $\rho(x,x,\omega=E_X)$.
    Two observations can be made.
    First, the spatial structure of the LDOS is faithfully reproduced by the tensor-network HODC calculation: the positions and relative weights of the localization peaks are highly correlated.
    Second, the KPM curves exhibit small negative oscillations, even at the level of the sparse KPM calculation. Although at $N_\mu=40$ the negative oscillations are inevitable, the KPM approximation can still resolve spatial features.
    In the large-scale calculations in \cref{fig: 1D 2D super moire}, similar negative oscillations occur while faithfully capturing the correct spatial correlations.

    \section{Discussion}
    The present results establish a scalable real-space framework for solving the BSE in regimes that are inaccessible to conventional approaches. In contrast to standard implementations, whose cost scales quadratically with the number of lattice sites and which require explicit storage of the two-particle Hamiltonian, our formulation exploits the TN structure to achieve logarithmic scaling with system size and efficient TN compression. This enables the treatment of super-moir\'e systems with effective Hamiltonian dimensions of order $10^{18}$, a regime far beyond existing methodologies. Working directly in real space is essential here: it lets the method treat aperiodic and incommensurate geometries, for which momentum-space BSE approaches tied to Bloch periodicity are not applicable, and thereby resolve multiscale excitonic phenomena within a single calculation, computations that are genuinely required for super-moiré periods of twisted van der Waals heterostructures that can span millions to billions of atoms.
    
    It is worth noting that several limitations remain. 
    In 2D, the rapid growth of bond dimensions with the number of Chebyshev moments places us near the threshold of efficient compressibility, currently limiting our accuracy to approximately forty moments. While the HODC kernel partially mitigates this by resolving spatial features of the LDOS at lower expansion orders, alternative architectures such as tree TNs \cite{Shi2006ClassicalNetwork} may be better suited to handle 2D correlations and further enhance performance. 
    A related constraint is the sensitivity of the bound-exciton signal to the SVD truncation tolerance: at large system sizes, an insufficiently tight tolerance selectively suppresses the bound-state spectral weight, so the tolerance must be chosen conservatively and tightened with system size (see the benchmarks in App.~\ref{App: extra benchmarks}).
    
    While we have focused here on local electron–hole interactions in minimal 1D and 2D lattices for concreteness, the framework can be systematically extended to incorporate long-range Coulomb interactions and exchange terms within the same TN formalism, as well as to more complicated lattice geometries. Such extensions would facilitate the quantitative modeling of realistic moiré heterostructures while preserving their real-space scalability demonstrated here. It is finally worth noting that the inclusion
    of long-range interactions and exchange terms may 
    potentially accelerate the growth of the bond dimension, 
    further motivating the development of alternative TN architectures.
    
    Beyond the specific exciton problem we have focused on here, our methodology provides a general strategy for addressing two-body correlation problems directly on real-space lattices \cite{Mattis1986TheLattice}. This technique can be applicable to other correlated states, including the formation of bipolaron states in moire systems \cite{Camacho-Guardian2018BipolaronsCondensate,Mazza2022StronglyHeterostructures,deSousaAraujoCassiano2024Strain-TuneableNanoribbon} and bimagnon bound states in moire magnetic insulators \cite{Song2021,Wong2026,Li2020MoireOrder,Ganguli2023VisualizationFerromagnet,2025arXiv250906753L,PhysRevB.103.L140409}. In all such cases, under appropriate circumstances, computations involve a two-particle Hilbert space whose direct treatment becomes prohibitive at large system sizes, yet can be efficiently encoded within the present TN framework.
        
	\section{Conclusion}
    % \JL{probably we need to fine tune this given the response to referees}
    % \YT{Maybe we can add also few sentences of the benchmarking/one two sentence condensing the discussion section}
    We have demonstrated a TN methodology that enables direct real-space calculations of bound-exciton spectral functions in exponentially large lattices. Our approach combines a pseudo-spin TN encoding of
    the BSE Hamiltonian, including both the single particle
    and many-body interaction in the full two-particle excitonic space. 
    Our algorithm leverages an interleaved representation of the electron–hole BSE space, together with a TN-based kernel polynomial expansion enhanced by a higher-order delta Chebyshev spectral kernel. 
    We demonstrated our methodology for 1D and 2D incommensurate super-moir\'e potentials with more than a billion sites, demonstrating the simultaneous resolution of atomistic and mesoscopic structures in the exciton LDOS. In both cases, we show the bound exciton spectral signatures in real space
    following the spatial modulation landscape. In 1D, we observed
    exciton miniband formation in real space and in 2D spatial confinement patterns at a specific exciton energy.
    Our methodology has been benchmarked against simple spinless two-band Wannierized Hamiltonians and shows good bond-dimension scaling for 1D, while 2D seems more limited at approximately $N_\mu=65$ with current architectures. Nevertheless, our approach provides a scalable real-space framework for exciton physics in moir\'e and super-moir\'e matter, offering a technique enabling the study of excitonic phenomena in super-moire materials.

	\begin{acknowledgments}
 
    We acknowledge the computational resources provided by the Aalto Science-IT project
    and the financial support from InstituteQ, 
    the
    Research Council of Finland (No. 370912), 
    the Finnish Ministry
    of Education and Culture through the Quantum Doctoral Education Pilot Program (QDOC VN/3137/2024-OKM-4), the
    Finnish Quantum Flagship (No. 358877, Aalto University),
    the Finnish Centre of Excellence in Quantum Materials QMAT
    (No. 374166),
    and the ERC Consolidator Grant ULTRATWISTROICS (No.
    101170477). LE acknowledges the research program “Materials for the Quantum Age” (QuMat) for financial support. 
    This program (registration number 024.005.006) is part of the Gravitation program financed by the Dutch Ministry of Education, Culture and Science (OCW).
    We thank P. Liljeroth, S. Kezilebieke, P. Shen,  E. Castro, B. Amorim, X. Waintal, A. Akhmerov, and C. M. Smith for useful discussions. 
    The code used for this work can be consulted at \cite{anouarrepo}.
    
	\end{acknowledgments}

\appendix

\begin{widetext}
\section{Derivation of the exciton Hamiltonian}\label{App: exciton Hamiltonian}

    We consider the set of two-particle states
    in which a single electron is excited from the filled valence band to the empty conduction band on a lattice of $ N$ sites.
    Before the excitation, the many-body ground state is given by $\ket{\Omega} = \prod_{l=1}^N c_{vl}^\dagger \ket{0}$, in which all valence orbitals are occupied, and all conduction orbitals are empty.
    The set of electron-hole excitations is given by $\ket{l,m} = c^{}_{vl} c^\dagger_{cm} \ket{\Omega}$, which describes a hole localized at the valence site $l$ and an electron at the conduction site $m$. The basis set $\{ \ket{l,m} \ | \ l,m=1,\cdots,N\}$ spans the exciton quasiparticle subspace, onto which we will project the full Hamiltonian.
    The Hamiltonian is described by the tight-binding model
    of Wannier conduction and valence states, 
    with 
    projected local inter-band interaction, taking the form
    \begin{equation}
    	\hat{H} = \sum_{b \in \{c,v\}}\sum_{lm} t_{lm}^b c^\dagger_{b l}c_{b m}^{} + \sum_l U_lc^\dagger_{vl}c^{}_{vl}c^\dagger_{cl}c^{}_{cl} \equiv H_T + H_U.
    \end{equation}
    To obtain the effective exciton Hamiltonian, we must calculate the matrix elements in the exciton basis, $H_{lk,l'k'} = \bra{l,k} \hat{H} \ket{l',k'}$. First, the kinetic contribution is
    \begin{equation*}
    \begin{aligned}
    	H^T_{lk,l'k'} &= \bra{l,k} \hat{H}_T \ket{l',k'}\\
    	&= \sum_{b \in \{c,v\}}\sum_{ij} t_{ij}^b \bra{\Omega} c_{ck}^{} c_{vl}^\dagger c^\dagger_{b i} c_{b j}^{} c_{vl'}^{} c_{ck'}^\dagger \ket{\Omega}\\
    	% &= \sum_{b \in \{c,v\}}\sum_{ij} t_{ij}^b \bra{\Omega} c_{vl}^\dagger c_{ck}^{} c^\dagger_{b i} c_{b j} c_{ck'}^\dagger c_{vl'}^{}  \ket{\Omega}\\
    	&= \sum_{b \in \{c,v\}}\sum_{ij} t_{ij}^b \bra{\Omega} c_{vl}^\dagger (\delta_{b c} \delta_{ki} - c^\dagger_{b i} c_{ck}^{}) (\delta_{b c} \delta_{k'j} - c_{ck'}^\dagger c_{b j}^{}) c_{vl'}^{}  \ket{\Omega}\\
    	&= \sum_{b \in \{c,v\}}\sum_{ij} t_{ij}^b \delta_{b c} \delta_{ki} \delta_{k'j} \bra{\Omega} c_{vl}^\dagger c_{vl'}^{} \ket{\Omega} + \sum_{b \in \{c,v\}}\sum_{ij} t_{ij}^b \bra{\Omega} c_{vl}^\dagger c^\dagger_{b i} c_{ck}^{} c_{ck'}^\dagger c_{b j}^{} c_{vl'}^{}\ket{\Omega}.
    \end{aligned}
    \end{equation*}
    In the last term, the cross-terms in the expansion vanish. This can be seen by commuting the conduction creation/annihilation operator and acting it on the Fermi sea. Upon further simplification,
    \begin{align*}
    	H^T_{lk,l'k'} &= \sum_{b \in \{c,v\}}\sum_{ij} t_{ij}^b \delta_{b c} \delta_{ki} \delta_{k'j} \bra{\Omega}\left( \delta_{ll'} - c_{vl'}^{} c_{vl}^\dagger \right)\ket{\Omega} + \sum_{b \in \{c,v\}}\sum_{ij} t_{ij}^b \bra{\Omega} c_{vl}^\dagger c^\dagger_{b i} (\delta_{kk'}-c_{ck'}^\dagger c_{ck}^{})  c_{b j}^{} c_{vl'}^{}\ket{\Omega}\\
    	% &= t_{kk'}^c \delta_{ll'} + \sum_{b \in \{c,v\}}\sum_{ij} t_{ij}^b \delta_{kk'} \bra{\Omega} c_{vl}^\dagger c^\dagger_{b i} c_{b j}^{} c_{vl'}^{}\ket{\Omega}\\
    	&= t_{kk'}^c \delta_{ll'} + \sum_{b \in \{c,v\}}\sum_{ij} t_{ij}^b \delta_{kk'} \bra{\Omega} c^\dagger_{b i} c_{vl}^\dagger  c_{vl'}^{} c_{b j}^{}\ket{\Omega}\\
    	% &= t_{kk'}^c \delta_{ll'} + \sum_{b \in \{c,v\}}\sum_{ij} t_{ij}^b \delta_{kk'} \bra{\Omega} c^\dagger_{b i} ( \delta_{ll'} - c_{vl'}^{} c_{vl}^\dagger) c_{b j}^{}\ket{\Omega}\\
    	&= t_{kk'}^c \delta_{ll'} + \sum_i t_{ii}^v \delta_{kk'} \delta_{ll'} - \sum_{b \in \{c,v\}}\sum_{ij} t_{ij}^b \delta_{kk'} \bra{\Omega} c^\dagger_{b i} c_{vl'}^{} c_{vl}^\dagger  c_{b j}^{}\ket{\Omega} \\
    	% &= t_{kk'}^c \delta_{ll'} + \sum_i t_{ii}^v \delta_{kk'} \delta_{ll'} - \sum_{b \in \{c,v\}}\sum_{ij} t_{ij}^b \delta_{kk'} \bra{\Omega} (\delta_{b v}\delta_{l'i} -  c_{vl'}^{} c^\dagger_{b i}) ( \delta_{b v} \delta_{lj}- c_{b j}^{} c_{vl}^\dagger )\ket{\Omega} \\
    	% &= t_{kk'}^c \delta_{ll'} + \sum_i t_{ii}^v \delta_{kk'} \delta_{ll'} - \sum_{b \in \{c,v\}}\sum_{ij} t_{ij}^b \delta_{kk'} \bra{\Omega} (\delta_{b v}\delta_{l'i} -  c_{vl'}^{} c^\dagger_{b i}) ( \delta_{b v} \delta_{lj}- c_{b j}^{} c_{vl}^\dagger )\ket{\Omega} \\
    	&= t_{kk'}^c \delta_{ll'} + \sum_i t_{ii}^v \delta_{kk'} \delta_{ll'} - \sum_{b \in \{c,v\}}\sum_{ij} t_{ij}^b \delta_{kk'} \delta_{b v}\delta_{l'i} \delta_{lj} \\
    	&= t_{kk'}^c \delta_{ll'} - t_{l'l}^v \delta_{kk'} + \sum_i t_{ii}^v \delta_{kk'} \delta_{ll'}.
    \end{align*}
    Here, the first term describes the hopping of the excited conduction electron. The second term corresponds to the hopping of the valence hole. The last term is a constant energy shift originating from the filled Fermi sea. It is diagonal in the exciton indices, such that it shifts all exciton energies and can be absorbed into a redefinition of the total energy.
    
    Now, for the interaction term, we have
    \begin{align*}
    	H^U_{lk,l'k'} &= \bra{l,k} \hat{H}_U \ket{l',k'}\\
    	&= \sum_i U_i \bra{\Omega}  c_{ck}^{} c_{vl}^\dagger c_{v_i}^\dagger c_{vi} c_{ci}^\dagger c_{ci}^{} c_{vl'}^{} c_{ck'}^\dagger \ket{\Omega}\\
    	% &= -\sum_i U_i \bra{\Omega} c_{ck}^{} c_{vl}^\dagger c_{v_i}^\dagger c_{vi} c_{ci}^\dagger  c_{vl'}^{} c_{ci} c_{ck'}^\dagger \ket{\Omega}\\
    	&= -\sum_i U_i\delta_{ik'} \bra{\Omega} c_{ck}^{} c_{vl}^\dagger c_{v_i}^\dagger c_{vi} c_{ci}^\dagger  c_{vl'}^{} \ket{\Omega}\\
    	% &= -\sum_i U_i\delta_{ik'} \bra{\Omega} c_{ck}^{} c_{vl}^\dagger (1-c_{vi} c_{v_i}^\dagger) c_{ci}^\dagger  c_{vl'}^{} \ket{\Omega}\\
    	&= -\sum_i U_i \delta_{ik'} \bra{\Omega} c_{ck}^{} c_{vl}^\dagger c_{ci}^\dagger  c_{vl'}^{} \ket{\Omega} + \sum_i U_i \delta_{ik'} \bra{\Omega} c_{ck}^{} c_{vl}^\dagger c_{vi} c_{vi}^\dagger c_{ci}^\dagger  c_{vl'}^{} \ket{\Omega}\\
    	% &= \sum_i U_i\delta_{ik'} \bra{\Omega} c_{ck}^{}  c_{ci}^\dagger c_{vl}^\dagger  c_{vl'}^{} \ket{\Omega} - \sum_i U_i\delta_{ik'} \bra{\Omega} c_{ck}^{}  (\delta_{il}-c_{vi}c_{vl}^\dagger)  c_{ci}^\dagger c_{vi}^\dagger  c_{vl'}^{} \ket{\Omega}\\\
    	% &= \sum_i U_i \delta_{ik'} \bra{\Omega}  (\delta_{ik} - c_{ci}^\dagger c_{ck}^{}) (\delta_{ll'} - c_{vl'}^{} c_{vl}^\dagger) \ket{\Omega} - \sum_i U_i\delta_{ik'} \delta_{il}\bra{\Omega} c_{ck}^{} c_{ci}^\dagger c_{vi}^\dagger  c_{vl'}^{} \ket{\Omega}\\
    	% &= \sum_i U_i\delta_{ik'} \delta_{ik} \delta_{ll'} - \sum_i U_i\delta_{ik'} \delta_{il}\bra{\Omega}  ( \delta_{ik}-c_{ci}^\dagger c_{ck}^{}) (\delta_{il'}-  c_{vl'}^{}c_{vi}^\dagger) \ket{\Omega}\\
    	&= \sum_i U_i\delta_{ik'} \delta_{ik} \delta_{ll'} - \sum_i U_i\delta_{ik'} \delta_{il} \delta_{ik} \delta_{il'}\\
    	&= U_k\delta_{kk'}\delta_{ll'} - U_k \delta_{kk'}\delta_{ll'}\delta_{kl}.
    \end{align*}
    The second term is the driving mechanism of bound exciton formation and is nonzero when $k=l$, i.e., when electron and hole occupy the same site, capturing the local electron-hole Coulomb interaction.
    
    To identify the physically relevant excitation energies, we subtract the ground-state energy $E_\Omega = \bra{\Omega} \hat{H} \ket{\Omega}$. First, we treat the kinetic term
    \begin{align*}
    	\bra{\Omega} \hat{H}_T \ket{\Omega} &= \sum_{b \in \{c,v\}}\sum_{ij} t_{ij}^b \bra{\Omega}c^\dagger_{b i}c_{b j}^{} \ket{\Omega}\\
    	&= \sum_{ij} t_{ij}^v \bra{\Omega}c^\dagger_{v i}c_{v j} \ket{\Omega}\\
    	% &= \sum_{ij} t_{ij}^v \bra{\Omega}(\delta_{ij}-c_{v j} c^\dagger_{v i})\ket{\Omega}\\
    	% &= \sum_{ij} t_{ij}^v \delta_{ij}\\
    	&= \sum_{i} t_{ii}^v.
    \end{align*}
    The interaction term, on the other hand is trivially $\bra{\Omega} \hat{H}_U\ket{\Omega}=0$. Thus $E_\Omega =\sum_{i} t_{ii}^v$.
    % \begin{align*}
    % 	\bra{\Omega} \hat{H}_U \ket{\Omega} &= \sum_i U_i\bra{\Omega} c_{vi}^\dagger c_{vi} c_{ci}^\dagger c_{ci} \ket{\Omega} = 0.
    % \end{align*}
    % From this we find
    % \begin{equation}
    % E_\Omega = \bra{\Omega} \hat{H}_T \ket{\Omega} + \bra{\Omega} \hat{H}_U \ket{\Omega} = \sum_{i} t_{ii}^v.
    % \end{equation}
    Since the filled valence sea contributes a constant background energy, we can subtract it from the effective exciton Hamiltonian to obtain
    \begin{align*}
    	H_{lk,l'k'} &= \bra{l,k} \hat{H} \ket{l',k'} - E_{\Omega} \delta_{kk'}\delta_{ll'}\\
    	&= \left(t_{kk'}^c + U_k\delta_{kk'} \right)\delta_{ll'} - t_{l'l}^v \delta_{kk'} - U_k \delta_{kk'}\delta_{ll'}\delta_{kl}.
    \end{align*}
    Written in operator form, this yields
    \begin{equation}\label{Eq: Exciton Hamiltonian}
        \hat{H}_X = \hat{T}_c\otimes\mathbbm{1} - \mathbbm{1}\otimes \hat{T}_v - \hat{U}_X
    \end{equation}
    where $\hat{T}_c$ contains both the conduction hopping term and an additional contribution from the interaction. The last term couples electrons and holes and cannot be written as a product operator.

\section{Construction of the interleaved-order MPO for general excitonic Hamiltonians}\label{App: interleaved order}
    
    We present the details of the construction of the Matrix Product Operator (MPO) employing an interleaved ordering strategy for general excitonic systems.
    While the Bethe-Salpeter Equation (BSE) serves as our primary illustrative example, this framework is inherently applicable to any
    $d$-dimensional tight-binding Hamiltonian where degrees of freedom can be partitioned into distinct sub-lattices or directions.
    
    First, we construct independent Hamiltonian MPOs for the electron ($e$) and hole ($h$) sectors, $\hat{\mathcal{H}}_e$ and $\hat{\mathcal{H}}_h$ respectively, using the standard site ordering as in Refs.~\cite{Sun2025Self-consistentSites,Antao2025TensorMosaics,Moustaj2026TensorSystems}. 
    For a system defined by $2^L$ spatial sites, these operators act separately on the local electron pseudo-spin basis $(s_{e_1}, s_{e_2}, \dots, s_{e_L})$ and hole pseudo-spin basis $(s_{h_1}, s_{h_2}, \dots, s_{h_L})$ respectively. 
    \begin{figure*}[t!]
        \centering
        \includegraphics{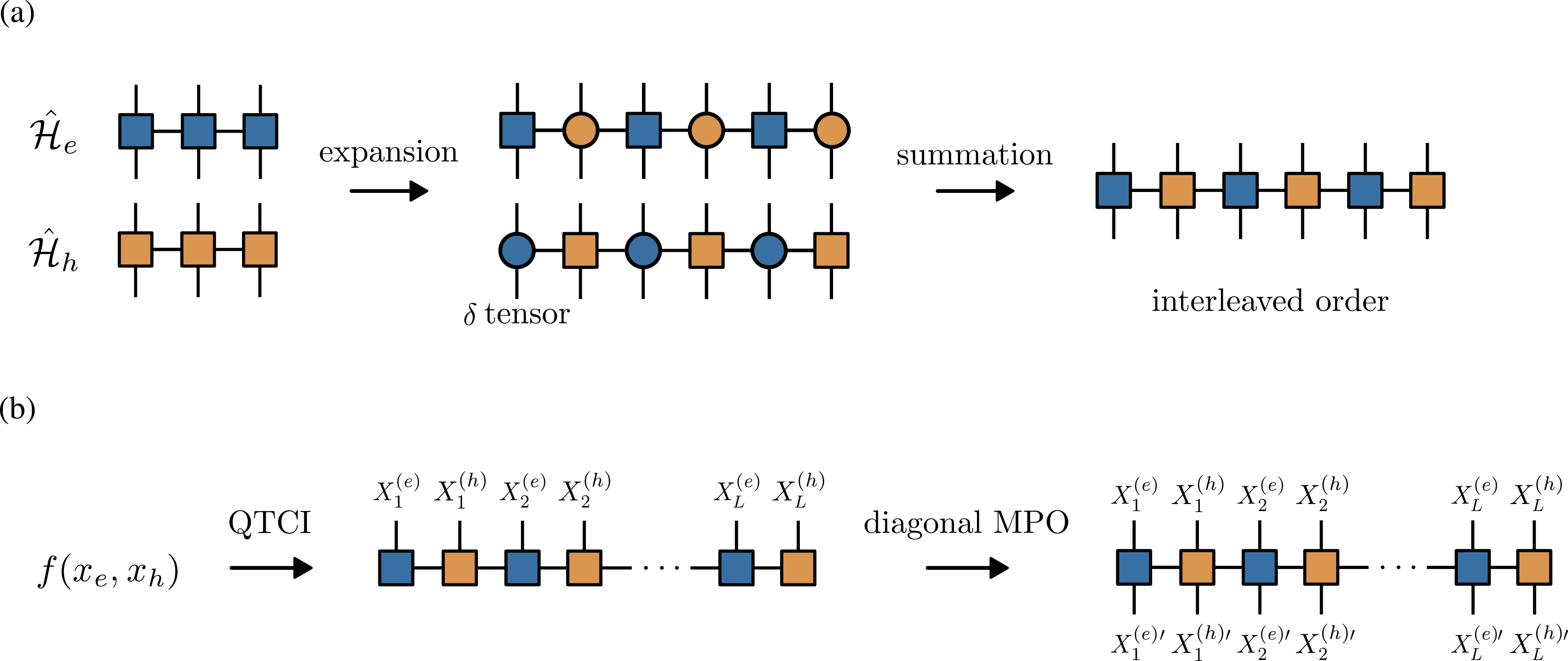}
        \caption{(a) The construction of the interleaved ordering for the Hamiltonian MPO. (b) An arbitrary function $f(x_e,x_h)$ of the electron and hole position is efficiently represented in interleaved order using QTCI and its promotion to a diagonal MPO.}
        \label{fig: interleaved order construction}
    \end{figure*}
    Subsequently, these individual MPOs are expanded into a unified $2L$-site Hilbert space through the insertion of $\delta$-tensors between adjacent blocks, as shown in \cref{fig: interleaved order construction}(a). There, we see that
    we implement a mapping where the electron degrees of freedom occupy the odd-numbered sites, and the hole degrees of freedom occupy the even-numbered sites.
    The $\delta$-tensors function as identity mappings that preserve the integrity of the electron and hole blocks, preventing unphysical coupling during the expansion. 
    By summing these expanded operators, we obtain the complete interleaved MPO, resulting in a basis of the form $(s_{e_1}, s_{h_1}, s_{e_2}, s_{h_2} \dots, s_{e_L},s_{h_L})$, and effectively placing corresponding electron and hole sites in immediate physical proximity along the chain. The resulting MPO is then simply $\hat{\mathcal{H}}_e\bigoplus\hat{\mathcal{H}}_h$ expressed in the interleaved order.
    
    % \begin{figure}[h!]
    %     \centering
    %     \includegraphics[width=0.9\linewidth]{SM/INTE.png}
    %     \caption{2D interpolation with QTCI by default is interleaved order.}
    %     \label{fig:2}
    % \end{figure}
     
    For the treatment of the correlated term $\hat{U}$ in \cref{Eq: Exciton Hamiltonian}, any spatial dependence of the form $U(\mathbf{r}_e, \mathbf{r}_h)$ can be directly implemented by
    learning the tensor network representaion with
    Quantics Tensor Train Cross Interpolation (QTCI) \cite{Ritter2024QuanticsFunctions,NunezFernandez2025LearningLibraries}. 
    Since the QTCI algorithm naturally performs multi-dimensional interpolation using a bit-interleaved order, 
    it will generate matrix product states (MPS) that are intrinsically compatible with our interleaved basis. 
    This eliminates the need for manual index remapping and ensures that interaction terms are treated optimally in the interleaved representation, as shown in \cref{fig: interleaved order construction}(b).
    
    The primary physical motivation for adopting this interleaved ordering lies in the preservation of operator locality and the subsequent reduction of the MPO bond dimension $\chi$. 
    As derived in the previous section, \cref{Eq: Exciton Hamiltonian} shows that for the highly localized excitonic system we are considering, the interaction term $U(\mathbf{r}_e, \mathbf{r}_h)$ only exists for electrons and holes at the same physical sites:
    \begin{equation}\label{Eq: onsite interaction}
        U(\mathbf{r}_e, \mathbf{r}_h) = \begin{cases} 
     
    U_0, & \mathbf{r}_e = \mathbf{r}_h \\
    0, & \mathbf{r}_e \neq \mathbf{r}_h 
    \end{cases}.
    \end{equation}
    In a minimal basis where all electrons precede all holes as $(s_{e_1}, s_{e_2}, \dots, s_{e_L}, s_{h_1}, s_{h_2} \dots,s_{h_L})$, 
    the MPO representation of \cref{Eq: onsite interaction} is $\hat{\mathcal{U}} = U_0\prod_{r=1}^L \left( \frac{1}{2} \mathbbm{1}_r \otimes \mathbbm{1}_{r+L} + 2  \sigma^{z}_r   \sigma^z_{r+L} \right)$. This operator necessitates capturing correlations between sites separated by a distance of $L$. Consequently, the MPO bond dimension $\chi$ scales as $2^L$ to account for these long-range entanglements.
    % \JL{is this necessarily true for a general non-interleaved order, or rather found empirically? meaning, is this true for an ordering Big1 Medium1 Small1 Small2 Medium2 Big2?}
    Conversely, in the interleaved basis $(s_{e_1}, s_{h_1}, s_{e_2}, s_{h_2} \dots, s_{e_L},s_{h_L})$, the same interaction becomes strictly local, involving only adjacent sites. 
    This locality allows the operator to be represented by a small and constant bond dimension that is independent of $L$, namely $\chi=2$ in our case. By distributing physical correlations more uniformly and locally across the tensor network, this ordering effectively mitigates the exponential growth of the bond dimension, thereby significantly enhancing both numerical precision and computational stability during large-scale simulations.
    
    Finally, it should be emphasized that the interleaved MPO is mathematically equivalent to the minimal basis discussed above, differing only by a permutation of the matrix elements. This representation preserves the Hamiltonian's spectrum while providing a vastly more efficient framework for tensor network calculations.

\begin{figure*}[!t]
    \centering
    \includegraphics[]{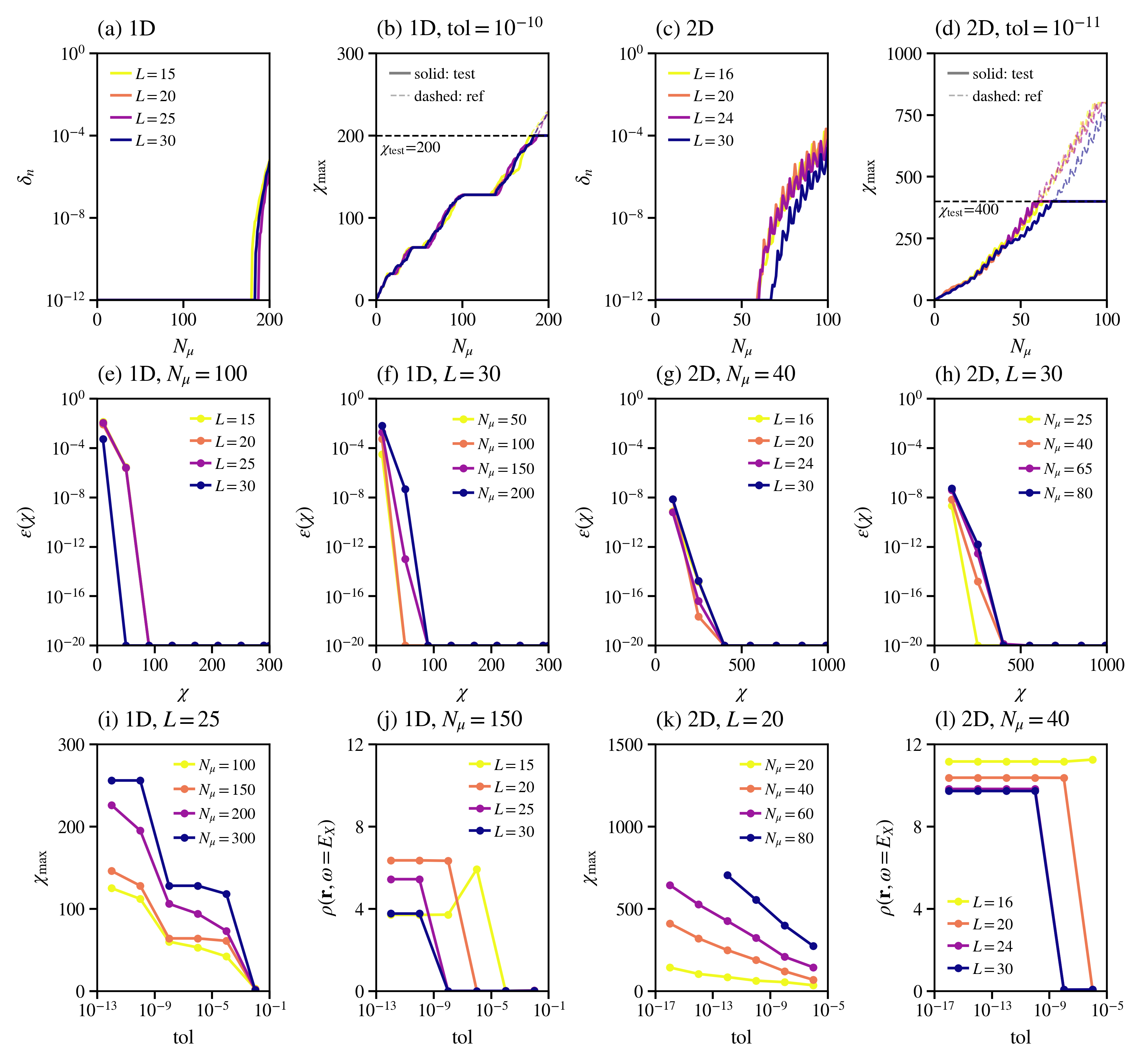}
    \caption{Benchmarking of the MPS exciton LDOS calculations. Panels (a--d) show Chebyshev-order convergence. Panels (a,c) show the infidelity $\delta_n$ between a test recursion with $\chi_\mathrm{test}=200$ and a reference recursion with $\chi_\mathrm{ref}=500$ as a function of Chebyshev order $N_\mu$, for the 1D and 2D models respectively. Panels (b,d) show the corresponding maximum bond dimension $\chi_\mathrm{max}$ for the test (solid) and reference (dashed) recursions; the horizontal dashed lines mark the $\chi_\mathrm{test}$ cap. Panels (e--h) show the normalized RMS LDOS error $\epsilon(\chi)$ versus KPM bond dimension $\chi$: panels (e,g) vary system size at fixed $N_\mu=100$ in 1D and $N_\mu=40$ in 2D, while panels (f,h) vary $N_\mu$ at fixed $L=30$ (in both 1D and 2D). Panels (i--l) show the dependence on compression tolerance $\mathrm{tol}$: panels (i,k) show the reached bond dimension $\chi_\mathrm{reached}$, and panels (j,l) show the bound-state LDOS signal $\rho(\mathbf{r},\omega=E_X)$. For the tolerance study, the fixed slices are $L=25$ in panel (i), $N_\mu=150$ in panel (j), $L=20$ in panel (k), and $N_\mu=40$ in panel (l). Colors encode increasing $L$ or increasing $N_\mu$ as indicated in each legend. System sizes are $N=2^L$, with $L\in{15,20,25,30}$ in 1D and $L\in{16,20,24,30}$ in 2D (square lattices, $L_x=L_y=L/2$)}
    \label{fig: full benchmark}
\end{figure*}

\section{Further Numerical benchmarks}\label{App: extra benchmarks}
    
    We first start by quantifying the accumulated truncation error by the infidelity between the two Chebyshev vectors at step $n$:
    \begin{equation}
    \delta_n = 1 -
    \frac{\bigl|\langle \nu_\text{ref}^n \mid \nu_\text{test}^n \rangle\bigr|^2}
    {\lVert\nu_\text{ref}^n\rVert^2\,\lVert\nu_\text{test}^n\rVert^2},
    \label{eq:infidelity}
    \end{equation}
    where $\ket{\nu_0} = \ket{\mathbf{r},\mathbf{r}}$, $\ket{\nu_1} = \tilde{H}\ket{\nu_0}$, $\ket{\nu_{n}} = 2\tilde{H}\ket{\nu_{n-1}} - \ket{\nu_{n-2}}$, and $\tilde{H}$ the rescaled Hamiltonian. A single probe site is evaluated per system: the deepest on-site potential well, i.e.\ the site $X=\arg\min_x V(x)$ at which the bound exciton has the largest $|X,X\rangle$ support. This is a consistent representative across system sizes, avoiding the size-dependent variability of sampling. In \cref{fig: full benchmark}, panels (a,c) show the infidelity $\delta_n$ between a test bond dimension and a reference bond dimension as a function of order $n$ in the Chebyshev recurrence for 1D ($\chi_\mathrm{test}=200$, $\chi_\mathrm{ref}=500$, tolerance $10^{-10}$) and 2D ($\chi_\mathrm{test}=400$, $\chi_\mathrm{ref}=800$, tolerance $10^{-11}$). Panels (b,d) show the corresponding maximum bond dimension $\chi_\mathrm{max}$ for the test (solid) and reference (dashed) recursions; the horizontal dashed line marks the $\chi_\mathrm{test}$ cap. System sizes are $N=2^L$, with $L\in{15,20,25,30}$ in 1D and $L\in{16,20,24,30}$ in 2D (square lattices, $L_x=L_y=L/2$). In 1D, the infidelity remains at the truncation floor up to $N_\mu\approx 180$ and stays below $\sim10^{-4}$ through $N_\mu=200$ across all sizes, while $\chi_\mathrm{max}$ grows steadily and the test recursion reaches its $\chi_\mathrm{test}=200$ cap near $N_\mu\approx 180$, the reference retaining headroom below $\chi_\mathrm{ref}=500$. In 2D the requirements are consistently more severe across sizes: $\chi_\mathrm{max}$ rises rapidly, the test recursion saturates its $\chi_\mathrm{test}=400$ cap by $N_\mu\approx 65$, and the reference approaches $\chi_\mathrm{ref}=800$ by $N_\mu=100$, indicating that the reference itself becomes bond-dimension-limited in this regime. The correspondingly low number of moments accessible in 2D is mitigated by the HODC kernel.
    
    To quantify the error in the reconstructed LDOS rather than in the propagated vectors, we compute the relative RMS error $\epsilon(\chi)$ [\cref{Eq: RMS error} in the main text] between the LDOS at bond dimension $\chi$ and a converged high-$\chi$ reference,
    \begin{equation}\label{Eq: app RMS error}
    \epsilon(\chi)= \sqrt{\frac{\sum_{ij}|\rho_{ij}(\chi)-\rho_{ij}(\chi_\text{ref})|^2}{\sum_{ij}|\rho_{ij}(\chi_\text{ref})|^2}},
    \end{equation}
    Panels (e–h) display this bond-dimension convergence: panels (e,g) fix the number of Chebyshev moments $N_\mu$ and vary the system size $L$ (1D, $N_\mu=100$, $L\in{15,20,25,30}$; 2D, $N_\mu=40$, $L\in{16,20,24,30}$), while panels (f,h) fix $L=30$ and vary $N_\mu$ (1D, $N_\mu\in{50,100,150,200}$; 2D, $N_\mu\in{25,40,65,80}$). In 1D (e,f) the error collapses to the numerical floor already at a modest bond dimension ($\chi\lesssim 90$), the required $\chi$ growing only mildly with $L$ and with $N_\mu$. In 2D (g,h) the decay is far slower and the error reaches the floor only near $\chi\sim 400$ (up to $\sim 500$ at the largest $N_\mu$), with the required $\chi$ increasing systematically with $N_\mu$; the largest moment count is the most demanding.
    
    Finally, panels (i–l) show the interplay between the SVD truncation tolerance (tol), the reached bond dimension, and the physical observable. Panels (i,k) show the maximum bond dimension $\chi_\mathrm{max}$ attained during the recursion as a function of tol, at fixed $L$ for several $N_\mu$: tightening the tolerance (moving left) monotonically increases $\chi_\mathrm{max}$, and the requirement grows with $N_\mu$, from $\chi_\mathrm{max}\sim 130$–$260$ in 1D (i, $L=25$) to $\chi_\mathrm{max}\sim 700$ in 2D (k, $L=20$, $N_\mu=80$); the 2D case also spans a substantially tighter tolerance window ($10^{-17}$–$10^{-5}$ versus $10^{-13}$–$10^{-1}$ in 1D). For $N_\mu=80$ in 2D, insufficient memory precluded calculations for $\mathrm{tol}<10^{-12}$. Panels (j,l) show the physical consequence: the bound-exciton spectral weight $\rho(\mathbf{r},\omega=E_X)$ at the exciton energy, evaluated at the deepest-well probe, as a function of tol (1D, $N_\mu=150$; 2D, $N_\mu=40$). The weight is essentially constant while the bound state is resolved but drops abruptly to zero once tol exceeds a size-dependent threshold: the $L=30$ signal is already lost near $\mathrm{tol}\sim 10^{-9}$, while smaller systems survive to looser tolerances. This makes it explicit that an insufficient truncation tolerance selectively destroys the bound-exciton signal and sets the required tolerance for each system size.
    
    Next, we compare the performance of the Jackson and HODC kernels in resolving the exciton LDOS at the modest number of moments $N_\mu=40$. In these calculations, we use $m=3$, which already provides improved resolution compared to the Jackson kernel, as shown in \cref{fig: kernel performance}. A higher value of $m$ while keeping $N_\mu=40$ will increase the negativity, as can be seen from \cref{Eq: Vandermonde system}, where $m$ enters as a power of the complex poles $z_l$ of \cref{Eq: rational approximant of delta}, therefore increasing the amplitude and frequency of oscillations. 
    For this reason, we do not consider it in this work. As $\eta$ decreases towards its ideal value of $1/N_\mu$, spectral broadening is reduced, and finer spatial features become more clearly resolved. This increased resolution, however, is accompanied by enhanced negative values in the density, arising both from tensor-network compression and from the intrinsic oscillatory nature of the HODC kernel. We note that the resutls in \cref{fig: kernel performance} were simulated using a higher compression rate than the figures of the main text. The tensor network parameters were set to $\chi_{\rm{max}} = 280$ and $\rm tol = 10^{-9}$.

    \begin{figure*}[!t]
        \centering
        \includegraphics[width=\textwidth]{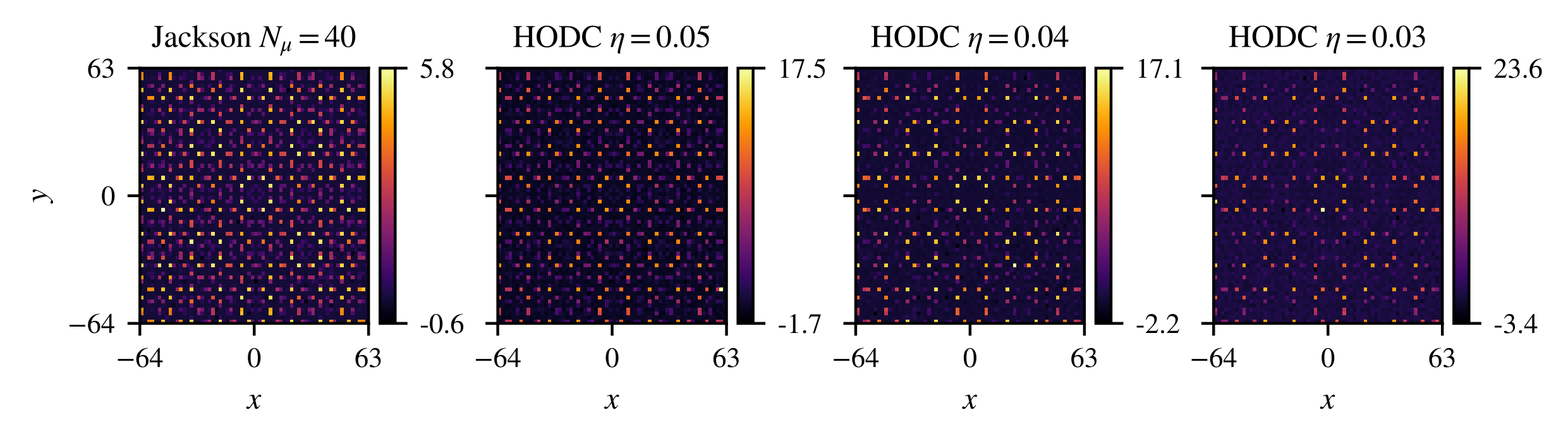}
        \caption{Comparison of the performance of the Jackson and HODC kernels at $N_\mu=40$ for the two-dimensional system considered in the main text, focusing on atomic-scale resolution. The row displays the raw results, where negative artifacts arise from tensor-network compression and, in the case of HODC, from the kernel's intrinsic properties. Note that these simulations were done at a higher compression rate than the high-resolution figures in the main text, as evidenced by the already present negativity of the LDOS for the Jackson kernel at $\rm min_\mathbf{r}\rho(\mathbf r,\omega)=-0.6$. In all cases, $N_\mu=40$ and $m=3$, $\chi_{\rm{max}}=280$ and $\rm tol = 10^{-9}$.}
        \label{fig: kernel performance}
    \end{figure*}

\end{widetext}

\bibliography{references_anouar,biblio_scf,biblio_chern_marker}
 % Produces the bibliography via BibTeX.

\end{document}